\documentclass[12pt]{article}
\usepackage{amsfonts}
\usepackage{epsfig}

\newcommand{\eq}{\begin{equation}}
\newcommand{\eqx}{\end{equation}}
\newcommand{\eqn}{\begin{eqnarray}}
\newcommand{\eqnx}{\end{eqnarray}}
\newcommand{\f}[2]{\frac{#1}{#2}}
\newcommand{\eps}{\varepsilon}
\newcommand{\tr}{\mbox{\rm tr}\,}
\newcommand{\res}{\mbox{\rm res}\,}

\newcommand{\Lm}{\Lambda}
\newcommand{\om}{\omega}
\newcommand{\gm}{\gamma}
\newcommand{\al}{\alpha}
\newcommand{\bt}{\beta}
\newcommand{\Dl}{\Delta}
\renewcommand{\th}{\theta}
\newcommand{\nn}{{\cal N}}
\newcommand{\CC}{{\cal C}}
\newcommand{\ZZ}{{\mathbb Z}}
\newcommand{\cor}[1]{\left\langle {#1} \right\rangle}
\newcommand{\qqqq}{\quad\quad\quad\quad}
\newcommand{\uf}{u^{fact.}}
\newcommand{\at}{\tilde{a}}
\newcommand{\taut}{\tilde{\tau}}

\title{Exact $U(N_c) \to U(N_1) \times U(N_2)$ factorization of
  Seiberg-Witten curves and $\nn=1$ vacua}

\author{Romuald A. Janik\footnote{e-mail: {\tt
ufrjanik@if.uj.edu.pl}}\\
M. Smoluchowski Institute  of Physics,\\ 
Jagellonian University,\\ 
Reymonta 4,\\
30-059 Krak\'{o}w, Poland
}

\begin{document}

\maketitle

\begin{abstract}
$\nn=2$ gauge theories broken down to $\nn=1$ by a tree level
superpotential are necessarily at the points in the moduli space where
the Seiberg-Witten curve factorizes. We find exact solution to the
factorization problem of Seiberg-Witten curves associated with the
breaking of the $U(N_c)$ gauge group down to two factors $U(N_1)\times
U(N_2)$. The result is a function of three discrete parameters and two
continuous ones. We find discrete identifications between various sets
of parameters and comment on their relation to the global structure of
$\nn=1$ vacua and their various possible dual descriptions. In an
appendix we show directly that integrality of periods leads to
factorization.  
\end{abstract}

\section{Introduction}

A breakthrough in our understanding of nonperturbative behaviour of
$\nn=2$ supersymmetric gauge theories occured in 1994 with the work
\cite{SW1,SW2} which showed that the low energy dyamics of the theory
was encoded in the properties of an associated hyperelliptic
Seiberg-Witten curve. 

Recently there has been much renewed interest in this circle of
problems associated with a set of completely new methods and insights
into the structure of $\nn=1$ theories obtained by
breaking $\nn=2$ theories to $\nn=1$ by an inclusion of a
superpotential for the adjoint $\nn=1$ chiral superfield. Calabi-Yau
constructions of the gauge theories, together with geometric
transitions \cite{V1,V2} gave rise to formulas for the exact
superpotential which where later reinterpreted in terms of a
computable random matrix theory \cite{DVP}. This correspondence 
was subsequently proven in a purely field-theoretic context
\cite{DVZANON,CDSW}. 

Since the $\nn=1$ theories which arise in the above way sit in the region
of $\nn=2$ moduli space where (some) monopoles become massless
i.e. where the Seiberg-Witten curves factorize this new circle of
ideas could be used as a new tool for studying the $\nn=2$ phenomena
and vice-versa. These interrelations were studied in
\cite{CV,FERRARI1,GOPAKUMAR,SCHNITZER1,SCHNITZER2,OBERS,deBOER}. In
\cite{DJ2} matrix model methods were used to obtain a solution of
the complete factorization of Seiberg-Witten curves for theories with
fundamental flavours (see also \cite{YD} for an application of these
results). 

The above dealt predominantly with the case when the gauge group was
not broken. In the opposite case e.g. for the breaking $U(N_c) \to
U(N_1)\times U(N_2)$ it was found \cite{PHASES1,FERRARI2} that the
space of $\nn=1$ vacua exhibits a very complex structure of various
connected components, each of which allows for {\em multiple} dual
descriptions of the same physics but with different patterns of
breaking. The analysis was later extended to the case of other gauge
groups and matter fields in the fundamental representation 
\cite{PHASES2,PHASESOTHER}.

However in the preceding in order to map out the structure of the
vacua one had to factorize the Seiberg-Witten curves for low $N_c$ on
a case by case basis. The aim of this paper is to provide an exact
solution of the factorization problem for arbitrary $N_c$ and any
pattern of breaking $U(N_c) \to U(N_1) \times U(N_2)$. 

The plan of this paper is as follows. In section 2 we give a short
introduction to the relation of factorized curves with $\nn=1$
vacua. In section 3 we summarize the equations for the meromorphic
1-form, the solution of which gives rise to the solution of the
factorization problem. In section 4, which is the main part of the
paper, we construct our solution. Then we proceed, in section 5, to
study its properties relevant for the uncovering of the global
structure of $\nn=1$ vacua. We close the paper with a
discussion and two appendices. In appendix A we give a mathematical
self-contained proof that factorization follows from integrality of
periods. We do this since the condition of integrality of periods as a
criterion for factorization was obtained using {\em gauge-theoretical}
considerations. In appendix B we list the {\it Mathematica} code which
implements our construction.   

\section{Factorization of Seiberg-Witten curves and $\nn=1$ vacua}

Since the celebrated work of \cite{SW1,SW2} it is known that $\nn=2$
$U(N_c)$ SYM theories develop an $N_c$-dimensional moduli space of
vacua parametrized by vacuum expectation values of the adjoint scalar
field
\eq
u_p =\cor{\f{1}{p} \tr \Phi^p}
\eqx
with $1 \leq p \leq N_c$.
The properties of the low energy dynamics of the theory around each
such vacuum is encoded in the geometry of the Seiberg-Witten curve
\cite{SW1,SW2,fa,Klemm} 
\eq
\label{e.swgen}
y^2=P_{N_c}(x,u_p)^2 -4 \Lm^{2N_c}
\eqx
where the polynomial $P_{N_c}(x,u_p)=\cor{\det(xI-\Phi)}\equiv
\sum_{\al=0}^{N_c} s_\al x^{N_c-\al}$ depends on
the moduli $u_p$'s parameterizing the vacua through
\eqn
&&\al s_\al +\sum_{k=0}^\al k s_{\al-k} u_k =0 \\
&&s_0=1, \qqqq u_0=0
\eqnx

The $\nn=2$ theory admits monopoles which at a generic point in the
moduli space are massive. However at special submanifolds of the
moduli space of vacua, $n$ of these monopoles ($n<N_c$) may become
massless and condense.  
At these points the Seiberg-Witten curve {\em factorizes} i.e. the rhs
of (\ref{e.swgen}) has $n$ double roots and $2(N_c-n)$ single roots:
\eq
\label{e.swfact}
P_{N_c} (x, \uf_p)^2 - 4 \Lambda^{2N_c} = F_{2(N_c-n)}(x)\cdot  H_n^2(x).
\eqx
The sets of moduli $\{ \uf_p \}$ then depend generically on $N_c-n$
{\em continuous} parameters and some discrete ones.

The submanifolds where the monopoles condense are also important due
to the fact that they are relevant when the $\nn=2$ theory is broken
down to $\nn=1$ through an addition of a tree level superpotential
\eq
W_{tree} = \sum_{p=1}^{n+1}  g_p \cdot \frac{1}{p} \tr \Phi^p.
\eqx 
Then the gauge symmetry $U(N_c)$ may be either unbroken, broken down
to $U(N_1)\times U(N_2)$ with $N_1+N_2=N_c$ etc. This is easy to
understand clasically as it corresponds to redistributing $N_c$
eigenvalues of $\Phi$ among respectively one, two or more minima of
the potential. On the quantum level the exact description is furnished
by the factorized form of the Seiberg-Witten curve (\ref{e.swfact})
with respectively $n=N_c-1$, $n=N_c-2$ etc.

The general solution for $n=N_c-1$ (complete factorization) has been
found in \cite{DS}. The solution is expressed in terms of
Chebyshev polynomials
\eq
\label{e.dssol}
P_{N_c} (x, \uf_p) = 2 \Lm^{N_c} \eta^{N_c} T_{N_c}\left(\f{x+x_0}{2\eta
    \Lm}\right) 
\eqx    
where $\eta=\exp(\pi i k/N_c)$.
Note that this solution depends both on a continuous parameter $x_0$
and on a discrete one $k=0... N_c-1$ (counting vacua in an $\nn=1$ $U(N_c)$
theory\footnote{The number of parameters is just $N_c$ since the
  solution depends only on $\eta^2$.}). 

Some solutions {\em for low} $N_c$ in the case of breaking $U(N_c) \to
U(N_1) \times U(N_2)$ were found on a case by case basis in
\cite{PHASES1}. The aim of this paper is to provide a general solution
in this case for any pattern of breaking characterized by $N_1$ and
$N_2$ and the appropriate discrete parameters $k_1$ and $k_2$.

Such a general solution is interesting especially as a very rich
structure of the vacua was uncovered for small $N_c$ in \cite{PHASES1}
(see also \cite{FERRARI2}).
In particular the space of vacua turns out to have multiple connected
components. Moreover within each component one may describe the same
physics (the same Seiberg-Witten curve) in terms of different labels
$(N_1,N_2,k_1,k_2)$ and $(N'_1,N'_2,k'_1,k'_2)$ corresponding to
different patterns of breaking of the gauge group. Note that one can
define the $N_i$'s and the appropriate glueball superfields $S_i$'s
not only in the semiclasssical limit ($\Lm \to 0$) but also
consistently at strong coupling through $N_i =(1/2\pi i) \oint_{\al_i}
\om $ where $\om$ is a meromorphic 1-form (see next section) while
$\al_i$ are some cycles. Dual descriptions correspond to different
choices of the fundamental cycles $\al_i$ but lead obviously to equivalent
physics. The main motivation for this work was the fact that the
knowledge of an exact solution of the factorization problem with
different patterns of breaking of the gauge group might help to map out a
global picture of the vacua and their possible dual descriptions.

\section{Integrality of periods}

A key object in the Seiberg-Witten theory is the meromorphic 1-form
\eq
\om=T(x) dx \equiv \cor{\tr \f{dx}{x-\Phi}}
\eqx
which is explicitly given for any Seiberg-Witten curve (\ref{e.swgen})
through \cite{CV,GOPAKUMAR,CDSW}
\eq
\om=T(x)dx = \f{P_{N_c}'(x) dx}{\sqrt{P_{N_c}^2(x)-4\Lm^{2N_c}}}
=\f{d}{dx} \log \left( P_N(x)+\sqrt{P_{N_c}^2(x)-4\Lm^{2N_c}} \right)
\eqx
From the definitions it is clear that 
\begin{enumerate}
\item  $T(x)$ has residue $N_c$ at infinity. 
\item The moduli of the vacuum $u_p$ can be reconstructed through
\eq
\label{e.upform}
u_p=\f{1}{p} \f{1}{2\pi i} \oint_{\CC_\infty} x^p \cdot \om =\f{1}{p}
\res_{x=\infty}\; x^p \cdot  \om
\eqx
\item The scale of the theory $\Lm$ can be reconstructed from the
  regularized integral
\eq
\left\{\int_a^\infty \om \right\}_{reg}  \equiv \lim_{x\to\infty}
\left(\int_a^x \om - N_c \log x \right)=-\log \Lm^{N_c}
\eqx
where $a$ is a branch point of the Seiberg-Witten curve.
\item If the Seiberg-Witten curve factorizes as in (\ref{e.swfact}),
  the 1-form $\om=T(x)dx$ defines a 1-form on the reduced curve
\eq
y^2=F_{2(N_c-n)}(x)
\eqx
\end{enumerate}

From the above properties we see that once we have found the meromorphic
1-form for the factorized curve we can reconstruct the moduli $u_p$
and thus the solution to the factorization problem. We need therefore
a criterion for the 1-form to come from a factorized curve. 

In \cite{PHASES2} and in \cite{FERRARI2} (using different notation --
see eqn. (A27) in the first paper of \cite{FERRARI2}) it was shown
from the Konishi anomaly 
perspective and the Dijkgraaf-Vafa superpotential respectively that
the periods of $\om=T(x)dx$ should be integer\footnote{We quote the
  equations for the case of $n=N_c-2$.}:
\eqn
\label{e.eone}
\f{1}{2\pi i} \oint_\al \om &=& N_1 \\
\label{e.etwo}
\f{1}{2\pi i} \oint_\bt \om &=& \Dl k \equiv k_1-k_2
\eqnx
The cycles $\al$ and $\bt$ in (\ref{e.eone})-(\ref{e.etwo}) are
ordinary {\em compact} cycles.
As mentioned above $\om$ can be understood to be a meromorphic 1-form
on the elliptic curve
\eq
\label{e.el}
y^2=F_4(x) \equiv (x-a)(x-b)(x-c)(x-d)
\eqx
The main aim of this paper is to give an explicit construction of the
general solution to these equations. 

Note that the equations (\ref{e.eone})-(\ref{e.etwo}) were obtained
from gauge-theoretical considerations. In appendix A we give a purely
mathematical proof which shows that factorization indeed follows from
the integrality of periods. That construction also shows a strong
structural similarity between the standard Chebyshev polynomials
appearing in (\ref{e.dssol}) and the polynomials constructed here
which solve the factorization problem. 

\section{Construction of the meromorphic 1-form}

Before we proceed to solve equations (\ref{e.eone})-(\ref{e.etwo}) let
us note some obvious symmetries of the problem. Firstly, given a solution
$P(x)$, the shifted polynomial $\tilde{P}(x)=P(x+x_0)$ is also
a solution. Secondly, a rescaling $x \to \al x$ corresponds
to a rescaling $\Lm^{2N_c} \to  \Lm^{2N_c}/\al^{2N_c}$. This allows us
to change $\Lm$ and also generate distinct solutions for fixed
$\Lm$ by picking $\al$ to be an appropriate root of unity.

The nontrivial part of the problem is then to find solutions
parametrized by $N_1$ and $\Dl k$ and one continuous parameter.

Let us note that in the original variables the problem seems
intractable. Namely we have to solve the equations
\eqn
\f{1}{\pi}\int_a^b \f{N_c x+ \mu}{\sqrt{(x-a)(x-b)(x-c)(x-d)}} dx &\sim&
N_1 \\
\f{1}{\pi}\int_b^c \f{N_c x+ \mu}{\sqrt{(x-a)(x-b)(x-c)(x-d)}} dx &\sim&
\Dl k
\eqnx
for two of the five unknowns $a,b,c,d$ and $\mu$. These are coupled
very nonlinear transcendental equations. And there seem to be no way
to obtain an explicit solution from the above expressions.

\subsubsection*{The torus with marked points}

The key to our solution is to use the representation of the elliptic
curve (\ref{e.el}) as a torus. However since the points at infinity
play a special role in our setup we have to pick two points $\tilde{a}_1$ and
$\tilde{a}_2$ on the torus (see fig. 1). These points will get mapped
to the $x=\infty$ infinities on the two branches of the curve (\ref{e.el}).

\begin{figure}[bt]
\centerline{%
\epsfysize=5cm 
\epsfbox{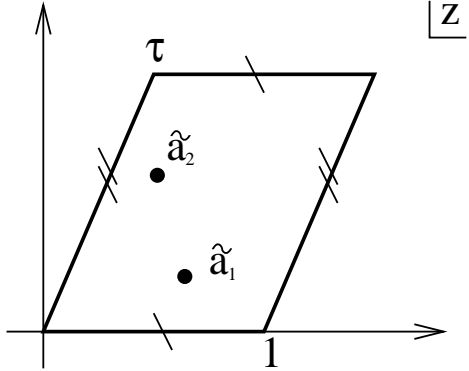}}
\caption{The torus with modular parameter $\tau$ and two marked points
  $\at_1$ and $\at_2$ which will be mapped to infinity.}
\end{figure}

The advantage of this representation is that holomorphic (and
meromorphic) 1-forms have a very simple form and periods may be
calculated quite explicitly. Indeed using the general representation of
a meromorphic function with prescribed poles and residues on a torus
\cite{MUMFORD} we may write $\om$ as
\eq
\om = \left( N_c \f{d}{dz} \log \f{\th(z-a_1)}{\th(z-a_2)} +C \right) dz
\eqx
where
\eq
\th(z)= \sum_{n=-\infty}^\infty e^{i \pi \tau n^2+2\pi i n z}
\eqx
is the standard theta function and $\tau$ is the modular parameter of
the torus. Then $\om$ has poles with residues $N_c$, $-N_c$ at
$\tilde{a}_1=\f{1+\tau}{2}+a_1$ and $\tilde{a}_2=\f{1+\tau}{2}+a_2$
respectively. 

Now it is trivial to calculate the periods. Using the periodicity
properties of the $\th$ function we get immediately
\eqn
\f{1}{2\pi i} \int_0^1 \om  &=& \f{C}{2\pi i} \\
\f{1}{2\pi i} \int_0^\tau \om  &=& N_c (a_1-a_2)+\f{C\tau}{2\pi i}
\eqnx
Of course due to the fact that $\om$ has poles with residue $\pm N_c$, the
above expressions are only defined up to a multiple of $N_c$.
We may now solve for the free parameters in $\om$ to get
\eqn
C &=& 2 \pi i N_1\\
a_2-a_1\equiv \Dl a &=& \f{N_1 \tau -\Dl k}{N_c}
\eqnx
In fact the final result will not depend on $a_1$ so we will set it to
zero.

\subsubsection*{Embedding of the torus}

In order to complete the solution of the factorization problem we need
to represent the torus with the modular parameter $\tau$ and the marked
points $a_1=0$ and $a_2=\Dl a$ as  the curve (\ref{e.el}). Once this is done
we will have at our disposal the function $x(z)$ which will allow us
to reconstruct the moduli parameters $u_p$ and the scale of the theory
$\Lm$. Comparing the general forms of the meromorphic 1-form in both
representations:
\eq
\f{N_c x +\mu}{y} dx = (N_c f(z)+C) dz
\eqx 
leads to the choice $x=f(z)$ and $y=dx/dz=f'(z)$. Although this choice
is not unique, we will use the freedom of rescaling $x \to \al x+x_0$
at the end of the construction. In principle the knowledge of $x(z)$
should be enough to calculate the $u_p$'s through (\ref{e.upform}),
however it is more convenient for numerical calculation to
reconstruct explicitly the curve (\ref{e.el}) and to calculate $u_p$'s
from 
\eq
\label{e.upres}
u_p = \f{1}{p} \res_{x=\infty}\;  x^p \cdot \f{N_c
  x+\mu}{\sqrt{(x-a)(x-b)(x-c)(x-d)}} 
\eqx

In order to reconstruct the equation satisfied by $y=f'(z)$ and
$x=f(z)$ one can use standard reasoning and look at the Laurent expansion
of $f(z)$ around it's pole at $z=\tilde{a}_1$:
\eq
f(z) \sim \f{1}{z-\at_1} + f_0 +f_1\, (z-\at_1) +f_2\, (z-\at_1)^2 +f_3\,
(z-\at_1)^3 
\eqx
Then one fixes the coefficients of the quartic polynomial (\ref{e.el}) by
requiring that in the Laurent expansion of the difference
\eq
\label{e.trial}
f'(z)^2 -(f^4(z) +S_1 f^3(z) +S_2 f^2(z) +S_3 f(z) +S_4)
\eqx
all poles in $(z-\at_1)$ and the constant term cancel. Then the
function (\ref{e.trial}) could have at most {\em one} pole at
$z=\at_2$ but on a torus that is impossible. Therefore (\ref{e.trial})
is holomorphic hence constant (and equal to zero). The elliptic curve
(\ref{e.el}) corresponding to our solution is then
\eq
y^2=x^4+ S_1 x^3 + S_2 x^2 +S_3 x + S_4
\eqx 
with the $S_i$'s given by
\eqn
S_1 &=& -4 f_0 \\
S_2 &=& 6f_0^2-6f_1 \\
S_3 &=& 12 f_0 f_1 -4 f_0^3 -8f_2 \\
S_4 &=& f_0^4+7f_1^2-6f_0^2 f_1 +8 f_0 f_2 -10 f_3
\eqnx
The parameter $\mu$ in (\ref{e.upres}) is just $\mu=2 \pi i N_1$.

\subsubsection*{The scale $\Lm$ of the theory}

It remains to evaluate the regularized integral related to $\Lm$. It
is convenient to calculate the equivalent quantity
\eq
-\log \Lm^{2N_c}=\lim_{\eps \to 0} \left[\int_{\at_1+\eps}^{\at_2-\eps} \om
+2\log \eps \right]
\eqx 
The result is
\eq
\log \Lm^2=2 \log \th'(\taut)-\log\left( \th(\taut-\Dl a)
  \th(\taut+\Dl a) \right) -\f{C \Dl a}{N_c} -i\pi
\eqx
where $\taut=(1+\tau)/2$.

This completes our solution. We have constructed a solution
parameterized by two discrete parameters $N_1$, $\Dl k$ and the
modular parameter of the elliptic curve $\tau$ in the upper half
plane. From this the standard rescaling $x \to \al x+x_0$ provides the
second continuous parameter $x_0$, allows us to fix $\Lm$ to the desired
physical value and introduces the third discrete parameter through the
appropriate root of unity in $\al$.

Explicitly we set
\eq
\label{e.al}
\al=-i\Lm^{-1} \cdot \th'(\taut)\left( \th(\taut-\Dl a)\th(\taut+\Dl a)
\right)^{-\f{1}{2}} e^{-\f{C \Dl a}{2N_c}} \cdot e^{-i\pi \f{k}{N_c}}
\eqx
where $\Lm$ is the physical scale of the theory, $C=2\pi i N_1$, $\Dl
a=(N_1\tau-\Dl k)/N_c$ and $k$ is the new discrete parameter. 

In appendix B we give the {\it Mathematica} code which implements the above
construction. 

\section{Some properties of the solution and the global structure of
  $\nn=1$ vacua}

The factorization solution constructed in the previous section
depends, for a given $N_c$, on the following set of parameters:
\eq
[N_1,k,\Dl k, \tau, x_0]
\eqx
where $N_1$, $k$ and $\Dl k$ are discrete, $\tau$ lies in the upper
half plane, while $x_0$ is an arbitrary complex number.

It may happen that two different sets of parameters describe the {\em
same} factorized Seiberg-Witten curve. This may happen for two
reasons. Firstly, as mentioned above the periods are defined only
modulo $N_c$, so we may expect that there are transformation rules
that allow us to reduce the parameters $\mbox{\sl mod } N_c$. Secondly one
expects e.g. that $\Dl k$ has simple periodicity w.r.t. $N_1$ and not
only with $N_c$. Thirdly, and
this is the most interesting possibility, the different sets of
(discrete) parameters for a single Seiberg-Witten curve represent
different dual descriptions of the same physics --- this is part of
the structure uncovered in \cite{PHASES1}. Indeed the description of the
identifications between the sets of discrete parameters describes
precisely the global structure of $\nn=1$ vacua.

If the Seiberg-Witten curve stays invariant, then necessarily the
elliptic curve also. Therefore identifications can only occur between
sets of parameters where $\tau$ and $\tau'$ are linked by a
$SL(2,\ZZ)$ transformation. We will first consider identifications for
fixed $\tau$ and later deal with the cases $\tau'=\tau+1$ and
$\tau'=-1/\tau$. 

\subsubsection*{Identifications with $\tau'=\tau$}

The most trivial property is periodicity in $k$ (see (\ref{e.al})):
\eq
\label{e.ei}
[N_1,k+2N_c,\Dl k, \tau, x_0]  \equiv  [N_1,k,\Dl k, \tau, x_0]
\eqx
Let us now consider a shift $\Dl k \to \Dl k+N_c$. Then $\Dl a \to \Dl
a-1$ and due to the invariance $\th(z)=\th(z+1)$ of the $\th$
functions the embdeding remains unchanged. The only modification is in
$\al$ which gets rescaled by a phase $e^{i\pi N_1/N_c}$. Hence we have
\eq
\label{e.eii}
[N_1,k,\Dl k+N_c, \tau, x_0]  \equiv  [N_1,k-N_1,\Dl k, \tau, x_0]
\eqx  
The shift $N_1 \to N_1+N_c$ is more complicated. Then $C\to C+2\pi i
N_c$, $\Dl a \to \Dl a+\tau$. Using the transformation law
$\th(z+\tau)=e^{-\pi \tau} e^{-2\pi i z} \th(z)$ we get $x \to x-2\pi
i$ and $\al \to \pm \al \cdot e^{-i\pi \Dl k/N_c}$. The sign
ambiguity is due to the branch cut of the square root in
(\ref{e.al}). Hence we get
\eq
\label{e.eiii}
[N_1+N_c,k,\Dl k, \tau, x_0] \equiv [N_1,k+\Dl k (+N_c),\Dl k, \tau,
x_0+2\pi i]  
\eqx

\subsubsection*{Identification with $\tau'=\tau+1$}

Under the modular transformation $\tau \to \tau+1$ the theta function
transforms as $\th(z,\tau+1)=\th(z+1/2,\tau)$. Therefore the elliptic curve
remains the same, the only thing that changes is the identification of
the periods namely
\eq
\Dl a=\f{N_1 (\tau+1)-\Dl k}{N_c} =\f{N_1 \tau -(\Dl k -N_1)}{N_c}
\eqx
Hence we have
\eq
\label{e.eiv}
[N_1,k,\Dl k, \tau+1, x_0] \equiv [N_1,k,\Dl k-N_1, \tau, x_0] 
\eqx
This transformation law has a clear physical meaning. In our
construction $\Dl k=k_1-k_2$ was periodic only modulo $N_c$, while we
know from the gauge theory perspective that $k_1$, which labels vacua
of $U(N_1)$ should have periodicity $N_1$. This identification is
furnished by the modular transformation $\tau \to \tau+1$.

\subsubsection*{Identification with $\tau'=-1/\tau$}

Here the modular transformation law of the theta function is more
involved:
\eq
\th\left(w, \f{-1}{\tau} \right) =(-i\tau)^{\f{1}{2}} e^{\pi i w^2
  \tau} \th(w \tau,\tau)
\eqx
The meromorphic function defining $x$ transforms as
\eq
\label{e.mers}
\f{d}{dw} \log \f{\th(w)}{\th(w-\Dl a')} \to \tau \left[ 2\pi i \Dl a'
  + \f{d}{dz} \log \f{\th(z)}{\th(z-\Dl a' \tau)} \right]
\eqx
where $z=w\tau$ and $\Dl a'=(-N'_1/\tau-\Dl k')/N_c$. The periods may
be calculated to be
\eq
N_1 =-\Dl k' \qqqq \Dl k= N'_1
\eqx
Finally $\al$ transforms simply as $\al \to \pm\tau \cdot \al$. This factor
exactly cancels the $\tau$ in (\ref{e.mers}) when constructing the
torus embedding. Putting all these ingredeints together we get
\eq
\label{e.ev}
\left[N_1,k,\Dl k, \f{-1}{\tau}, x_0 \right] \equiv \left[-\Dl
  k,k(+N_c),N_1, \tau, \f{x_0}{\tau}+\f{2\pi i}{N_c}\left(
  \f{N_1}{\tau}+\Dl k \right) \right]
\eqx  

This is the key transformation which allows us to obtain a dual
description with a different pattern of breaking $U(N_c) \to U(N_1)
\times U(N_2)$. 

\subsubsection*{Global structure of vacua}

The above identifications allow us to study the global structure of
$\nn=1$ vacua. The connected components could be obtained by finding
the orbits of the {\em three} discrete labels $[N_1,k,\Dl k]$ under
the group of transformations generated by (\ref{e.ei}), (\ref{e.eii}),
(\ref{e.eiii}), (\ref{e.eiv}) and (\ref{e.ev}), assuming that there
are no further identifications.  
We leave the detailed investigation of this structure to a subsequent
work \cite{WIP}.

\section{Discussion}

In this paper we constructed an exact solution of the factorization
problem of Seiberg-Witten curves with $n=N_c-2$ massless
monopoles. This is relevant for the breaking of $\nn=2$ theories down
to $\nn=1$ with the gauge group $U(N_c)$ broken down to a product of
two factors. 

The solution was obtained by constructing a meromorphic 1-form with
integral periods on an elliptic curve. The solution depends on three
discrete parameters and two continuous ones.
In principle one could
generalize the structure to lower $n$ by considering the construction
of meromorphic functions on hyperelliptic curves of higher genus. 
Another direction in generalization could be the extension to other
gauge groups and/or addition of matter fields.
A more complicated but interesting setting would be the consideration
of quiver gauge theories, where nonhyperelliptic curves appear, and
where it was shown that integrality of periods also plays an important
role \cite{QUIVER}.

We also studied certain discrete identifications between the
parameters and in appendix A we gave a proof that integrality
of periods, a condition obtained in \cite{PHASES2,FERRARI2} on
physical grounds, indeed leads to a solution of the factorization
problem. 

The physical interest of an exact solution lies in the
possibility of studying in detail the global structure of $\nn=1$
vacua for any $N_c$ along the lines of \cite{PHASES1}. The main
features of the analysis in \cite{PHASES1}, like the appearance of
connected components of vacua and possible dual descriptions are
related to discrete identifications between the parameters labeling
the factorization solution. We leave a detailed investigation of this
question for future work \cite{WIP}.

\bigskip

\noindent{\bf Acknowledgments} RJ would like to thank 
the Niels Bohr Institute for hospitality while this work was carried
out. This work was supported in part by KBN grants~2P03B09622
(2002-2004), 2P03B08225 (2003-2006) and by ``MaPhySto'', the Center of
Mathematical Physics and Stochastics financed by the National Danish
Research Foundation. 

\medskip

\appendix

\section*{Appendix A. A mathematical proof of the factorization property of our
  solution}

The main object of this paper was to construct a solution of the
factorization problem by constructing explicitly a meromorphic 1-form
$\om$ on an elliptic curve
\eq
\label{e.elcurve}
y^2=(x-a)(x-b)(x-c)(x-d)
\eqx
with the following properties:
\eqn
\label{e.intper}
&&\f{1}{2\pi i} \oint_{\gm_i} \om \in \ZZ \\
\label{e.resid}
&&\res_{x=\pm \infty}\; \om = \pm N_c
\eqnx
These properties followed from the expressions for the superpotential
in the Dijkgraaf-Vafa framework. Then we reconstructed the parameters
of the factorized curve from the `physical' formulas
\eq
u_p = \f{1}{p}\res_{x=\infty}\; x^p \cdot \om
\eqx

In this appendix we show directly, without any physical input, that
the conditions (\ref{e.intper})-(\ref{e.resid}) give rise to a
solution of the factorization problem, namely they allow us to
construct a polynomial $P(x)=x^{N_c}+\ldots$ such that the
Seiberg-Witten curve factorizes:
\eq
P^2-1=  (x-a)(x-b)(x-c)(x-d)  \cdot H(x)^2
\eqx
with $H(x)$ a polynomial of degree $N_c-2$. From the construction in
the main text we see that we may assume without loss of generality that 
\eq
\label{e.norm}
\left\{\int_a^\infty \om \right\}_{reg}  \equiv \lim_{x\to\infty}
\left(\int_a^x \om - N_c \log x \right) =\log 2
\eqx

Let us define the holomorphic function $P(x)$ on the complex plane outside the
cuts defined by (\ref{e.elcurve}) through
\eq
\label{e.pdef}
P(x)=\cosh \left( \int_a^x \om \right)
\eqx
where $a$ is a branching point of the curve (double covering)
(\ref{e.elcurve}).
Since the periods of $\om$ are integer this is a well-posed definition
which does not depend on the path of integration. 

One can then show that $P(x)$ (and its derivative) is continuous across
the cuts. This follows from the fact that across the cut $\om$ changes
just by a sign\footnote{And consequently one can show that $P$ is $\pm
  1$ at the branch points of (\ref{e.elcurve}).}. Hence $P(x)$ defined
by (\ref{e.pdef}) can be extended to an entire function on the complex
plane.

Then the requirement that the residue of $\om$ is $N_c$ and the
normalization integral (\ref{e.norm}) leads to the asymptotic behaviour
\eq
\int_a^x \om \sim N_c \log x + \log 2 + \ldots
\eqx
which shows that $P(x)$ must be a polynomial of order $N_c$ with unit
coefficient. 

We are now ready to show that factorization occurs. Let us denote
$F_4=(x-a)(x-b)(x-c)(x-d)$. Then the preceding footnote shows that we
have
\eq
P^2-1=F_4 \cdot G
\eqx
We have to show that $G$ has zeroes with even multiplicity. To this
end consider the 1-form 
\eq
\sqrt{F_4} \cdot  \f{dP}{\sqrt{P^2-1}} =\f{P'(x) dx}{\sqrt{G}}
\eqx
but this is exactly equal to
\eq
\sqrt{F_4} \cdot \f{\om \sinh \int_a^x \om}{ \sinh \int_a^x \om}
=\sqrt{F_4} \om
\eqx
which is a polynomial 1-form. Hence the roots of $G$ have to appear in
pairs which proves factorization
\eq
P^2-1=F_4 \cdot H^2
\eqx

Note that the above reasoning works for any number of cuts and hence
for any meromorphic 1-form with integer periods and fixed residues at
infinity on an hyperelliptic curve.

It is amusing to see that the same construction generates the classical
Chebyshev polynomials appearing in the solution of the complete
factorization problem by Douglas and Shenker \cite{DS}. There, the relevant
formula would be\footnote{Here for simplicity we relaxed the normalization
  constraint (\ref{e.norm}).} 
\eq
P_{N_c}(x)=\cosh \left( \int_1^x \f{N_c dx}{\sqrt{x^2-1}} \right)
\eqx
The above formula indeed gives exactly the Chebyshev polynomial
$T_{N_c}(x)$. Our formula (\ref{e.pdef}) thus gives a very natural
generalization of Chebyshev polynomials from the sphere to elliptic
and hyperelliptic curves.

\section*{Appendix B. {\it Mathematica} code which implements the
  construction of the factorized Seiberg-Witten curve}

In this appendix we enclose the {\it Mathematica} code for generating
the factorized Seiberg-Witten curve. The calling sequence is\\ 
{\tt generate[nc,n1,k,dk,tau,x0,dl]}\\
where {\tt dl} stands for $\Lm^{2N_c}$. All calculations have to be
performed with high numerical accuracy.

\begin{verbatim}
prec=30  (* numerical precision *)

mer[a_,z_]:=Pi*EllipticThetaPrime[3,Pi(z-a),q]/EllipticTheta[3, Pi(z-a),q]

(* torus embedding *)

setpa:=({pa,pb,pc,pd,pe} =
        {(D[EllipticTheta[3, Pi(z-a1),q],z] /. z->tt+a1),
        ((1/2)D[EllipticTheta[3, Pi(z-a1),q],z,z] /. z->tt+a1),
        ((1/6)D[EllipticTheta[3, Pi(z-a1),q],z,z,z] /. z->tt+a1),
        ((1/24)D[EllipticTheta[3, Pi(z-a1),q],z,z,z,z] /. z->tt+a1),
        ((1/120)D[EllipticTheta[3, Pi(z-a1),q],z,z,z,z,z] /. z->tt+a1)})

setaa:=({aa,bb,cc,dd} ={ pb/pa, -((pb^2 - 2*pa*pc)/pa^2),
        (pb^3 - 3*pa*pb*pc + 3*pa^2*pd)/pa^3,
        -((pb^4 - 4*pa*pb^2*pc + 4*pa^2*pb*pd + 
        2*pa^2*(pc^2 - 2*pa*pe))/pa^4) } )

seta:=({f0,f1,f2,f3} =
        { (aa-mer[a2,z]) /. z->tt+a1,
        (bb-D[mer[a2,z],z]) /. z->tt+a1,
        (cc-(1/2)D[mer[a2,z],z,z]) /. z->tt+a1,
        (dd-(1/6)D[mer[a2,z],z,z,z]) /. z->tt+a1})

sets:= ({s1,s2,s3,s4} = {-4f0, 6(f0^2-f1), 12 f0 f1 -4f0^3-8f2, 
        f0^4+7f1^2-6f0^2 f1+8f0 f2-10 f3})

defcurve:=setpa; setaa; seta; sets

(* Seiberg-Witten curves *) 

u[i_]:=data[[i]]

s[0]=1
s[r_]:=N[-Sum[ s[r-a] a u[a], {a,1,r} ]/r,prec]

swpol[nc_]:=Sum[ s[a] x^(nc-a), {a,0,nc} ]

swc[nc_,dl_]:=(swpol[nc]^2-4 dl)

(* main procedure *)

generate[nc_,n1_,k_,dk_,tau_,x0_,dl_]:=Block[{},
  nu=n1/nc; deltak=dk/nc;
  a1=0; 
  a2=nu tau-deltak; 
  q=Exp[Pi I tau]; 
  tt=(1+tau)/2;
  defcurve;
  Print["Generating..."];
  thetafncs:=-I*Pi EllipticThetaPrime[3,Pi tt,q] *
   (EllipticTheta[3,Pi(a2-a1+tt),q]*EllipticTheta[3,Pi(a1-a2+tt),q])^(-1/2)*
    Exp[-2Pi I n1 (n1 tau-dk)/(2*nc^2)];
  al=dl^(-1/(2nc)) thetafncs Exp[-I*Pi*k/nc];
  forminit=(x+2 Pi I nu)/Sqrt[x^4+s1 x^3+ s2 x^2+s3 x +s4];
  form=N[al*forminit /. x->al*x+x0, prec];
  data=Table[-nc Residue[form*x^i,{x,Infinity}]/i, {i,1,nc}];
  data=data/(-Residue[form,{x,Infinity}]);
  Chop[swc[nc,dl]]]
\end{verbatim}


\begin{thebibliography}{99}

\bibitem{SW1}
N.~Seiberg and E.~Witten,
``Electric - magnetic duality, monopole condensation, and confinement
in N=2 supersymmetric Yang-Mills theory,'' 
Nucl.\ Phys.\ B {\bf 426} (1994) 19
[Erratum-ibid.\ B {\bf 430} (1994) 485]
[arXiv:hep-th/9407087].

\bibitem{SW2}
N.~Seiberg and E.~Witten,
``Monopoles, duality and chiral symmetry breaking in N=2 supersymmetric QCD,''
Nucl.\ Phys.\ B {\bf 431} (1994) 484
[arXiv:hep-th/9408099].

\bibitem{V1}
C.~Vafa,
``Superstrings and topological strings at large N,''
J.\ Math.\ Phys.\  {\bf 42} (2001) 2798
[arXiv:hep-th/0008142].

\bibitem{V2}
F.~Cachazo, K.~A.~Intriligator and C.~Vafa,
``A large N duality via a geometric transition,''
Nucl.\ Phys.\ B {\bf 603} (2001) 3
[arXiv:hep-th/0103067].

\bibitem{DVP}
R.~Dijkgraaf and C.~Vafa,
``A perturbative window into non-perturbative physics,''
arXiv:hep-th/0208048.

\bibitem{DVZANON}
R.~Dijkgraaf, M.~T.~Grisaru, C.~S.~Lam, C.~Vafa and D.~Zanon,
``Perturbative computation of glueball superpotentials,''
Phys.\ Lett.\ B {\bf 573} (2003) 138
[arXiv:hep-th/0211017].

\bibitem{CDSW}
F.~Cachazo, M.~R.~Douglas, N.~Seiberg and E.~Witten,
``Chiral rings and anomalies in supersymmetric gauge theory,''
JHEP {\bf 0212} (2002) 071
[arXiv:hep-th/0211170].

\bibitem{CV}
F.~Cachazo and C.~Vafa,
``N = 1 and N = 2 geometry from fluxes,''
arXiv:hep-th/0206017.

\bibitem{FERRARI1}
F.~Ferrari,
``On exact superpotentials in confining vacua,''
Nucl.\ Phys.\ B {\bf 648} (2003) 161
[arXiv:hep-th/0210135].

\bibitem{GOPAKUMAR}
R.~Gopakumar,
``N = 1 theories and a geometric master field,''
JHEP {\bf 0305} (2003) 033
[arXiv:hep-th/0211100].

\bibitem{SCHNITZER1}
S.~G.~Naculich, H.~J.~Schnitzer and N.~Wyllard,
``The N = 2 U(N) gauge theory prepotential and periods from a
perturbative matrix model calculation,'' 
Nucl.\ Phys.\ B {\bf 651} (2003) 106
[arXiv:hep-th/0211123].

\bibitem{SCHNITZER2}
S.~G.~Naculich, H.~J.~Schnitzer and N.~Wyllard,
``Matrix model approach to the N = 2 U(N) gauge theory with matter in
the  fundamental representation,'' 
JHEP {\bf 0301} (2003) 015
[arXiv:hep-th/0211254].

\bibitem{OBERS}
R.~A.~Janik and N.~A.~Obers,
``SO(N) superpotential, Seiberg-Witten curves and loop equations,''
Phys.\ Lett.\ B {\bf 553} (2003) 309
[arXiv:hep-th/0212069].

\bibitem{deBOER}
V.~Balasubramanian, J.~de Boer, B.~Feng, Y.~H.~He, M.~x.~Huang,
V.~Jejjala and A.~Naqvi, 
``Multi-trace superpotentials vs. matrix models,''
Commun.\ Math.\ Phys.\  {\bf 242} (2003) 361
[arXiv:hep-th/0212082].

\bibitem{DJ2}
Y.~Demasure and R.~A.~Janik,
``Explicit factorization of Seiberg-Witten curves with matter from
random  matrix models,''
Nucl.\ Phys.\ B {\bf 661} (2003) 153
[arXiv:hep-th/0212212].

\bibitem{YD}
Y.~Demasure,
``Affleck-Dine-Seiberg from Seiberg-Witten,''
arXiv:hep-th/0307082.

\bibitem{PHASES1}
F.~Cachazo, N.~Seiberg and E.~Witten,
``Phases of N = 1 supersymmetric gauge theories and matrices,''
JHEP {\bf 0302} (2003) 042
[arXiv:hep-th/0301006].

\bibitem{FERRARI2}
F.~Ferrari,
``Quantum parameter space and double scaling limits in N = 1 super
Yang-Mills theory,'' 
Phys.\ Rev.\ D {\bf 67} (2003) 085013
[arXiv:hep-th/0211069];\\
F.~Ferrari,
``Quantum parameter space in super Yang-Mills. II,''
Phys.\ Lett.\ B {\bf 557} (2003) 290
[arXiv:hep-th/0301157].

\bibitem{PHASES2}
F.~Cachazo, N.~Seiberg and E.~Witten,
``Chiral Rings and Phases of Supersymmetric Gauge Theories,''
JHEP {\bf 0304} (2003) 018
[arXiv:hep-th/0303207].

\bibitem{PHASESOTHER} 
C.~h.~Ahn and Y.~Ookouchi,
``Phases of N = 1 supersymmetric SO / Sp gauge theories via matrix model,''
JHEP {\bf 0303} (2003) 010
[arXiv:hep-th/0302150];
V.~Balasubramanian, B.~Feng, M.~x.~Huang and A.~Naqvi,
``Phases of N = 1 supersymmetric gauge theories with flavors,''
arXiv:hep-th/0303065;
C.~h.~Ahn, B.~Feng and Y.~Ookouchi,
``Phases of N = 1 SO(N(c)) gauge theories with flavors,''
arXiv:hep-th/0306068;
P.~Merlatti,
``Gaugino condensate and phases of N = 1 super Yang-Mills theories,''
arXiv:hep-th/0307115;
C.~h.~Ahn, B.~Feng and Y.~Ookouchi,
``Phases of N = 1 USp(2N(c)) gauge theories with flavors,''
arXiv:hep-th/0307190.

\bibitem{fa}
P.~C.~Argyres and A.~E.~Faraggi,
``The vacuum structure and spectrum of N=2 supersymmetric SU(n) gauge
theory,'' 
Phys.\ Rev.\ Lett.\  {\bf 74} (1995) 3931
[arXiv:hep-th/9411057].

\bibitem{Klemm}
A.~Klemm, W.~Lerche, S.~Yankielowicz and S.~Theisen,
``Simple singularities and N=2 supersymmetric Yang-Mills theory,''
Phys.\ Lett.\ B {\bf 344} (1995) 169
[arXiv:hep-th/9411048].

\bibitem{DS} 
M.~R.~Douglas and S.~H.~Shenker,
``Dynamics of SU(N) supersymmetric gauge theory,''
Nucl.\ Phys.\ B {\bf 447} (1995) 271
[arXiv:hep-th/9503163].

\bibitem{MUMFORD} D. Mumford, ``Tata lectures on theta. I'', Progress
  in Mathematics 28, Birkhauser Boston, 1983.

\bibitem{WIP} Work in progress.

\bibitem{QUIVER}
R.~Casero and E.~Trincherini,
``Phases and geometry of the N = 1 A(2) quiver gauge theory and
matrix models,'' 
JHEP {\bf 0309} (2003) 063
[arXiv:hep-th/0307054].

\end{thebibliography}
\end{document}